\documentclass[pre,twocolumn,showpacs,preprintnumbers,amsmath,amssymb]{revtex4} 
\usepackage{amsmath}
\usepackage{graphicx}
\usepackage{dcolumn}
\usepackage{bm}
\input epsf

\begin{document}
\title{Population Dynamics in the Penna Model}
\author{J.~B.~Coe and Y.~Mao}
\affiliation{Cavendish Laboratory, Madingley Road, Cambridge, CB3 OHE, United Kingdom}


\begin{abstract}
We build upon the recent steady-state Penna model solution, {\it Phys.\ Rev.\ Lett.\ 89, 288103 (2002)},
to study the population dynamics within the Penna model. We show, that any perturbation to the 
population can be broken into a collection of modes each of which decay exponentially with its respective 
time constant. The long time behaviour of population is therefore likely to be dominated by the 
modes with the largest time constants. We confirm our analytical approach with simulation data.
\end{abstract}

\maketitle

\section{Introduction}

Genes, mutation, evolution and ageing have been topics of intensive research \cite{book1,book2,book3,book4,jaz},
particularly after the recent Genome project. In 1995, a bit-string computer simulation model for 
population evolution was introduced by Penna \cite{Penna} which successfully encompassed all those elements.
The Penna model, essentially a mutation accumulation model, has been so successful that it has rapidly 
established itself as a major model for population simulations \cite{book4}. Recently, analytical
solutions have been presented for the steady states of the Penna model \cite{Coe_Mao_Cates,Analytical_solution}, 
shedding insights into the inter-relationships between various Penna parameters of the simulation model.
Here, we build upon this analytical framework to study populations
undergoing steady growth or decline and to study the transient behaviour of the model
when the system is away from its steady states. 
Previous attempts \cite{piza} to analyse the population dynamics did not have the advantage of the full analytic
steady state solution \cite{Coe_Mao_Cates,Analytical_solution}, and consequently could not fully explore the
dynamics of the model. We find that the 
fluctuations in a population away from steady state can be decomposed into a collection of 
modes each of which decays 
exponentially with its respective time constant. The long time behaviour of a population will therefore be
dominated by the modes with the largest time constants. Our analytical results are confirmed by comparison
with simulation results.

\section{The Simple Penna Model}

In the Penna model \cite{Penna}, an organism's genome is represented
by a bit-string. Organisms age in time steps and at each timestep an organism reads the corresponding
bit from the bit string (e.g.~2nd site at age 2). If a site contains a $1$ the organism develops a disease,
and once it has accumulated $T$ diseases it dies. In any timestep an organism can reproduce with probability
$b$. The child's bit-string is a copy of the parent's with a probability $m$ of each site mutating
into a $1$. Positive mutations are rare in nature so a $1$ mutating into a $0$ is forbidden in the model.
Variants of the Penna model exist in which an organism can only reproduce between certain ages, and in which
there is an external death rate giving each organism a genome-independent chance of death in any timestep.

Traditionally \cite{Stauffer_MC} the Penna model is implemented computationally and the population controlled by the 
use of a Verhulst factor \cite{Verhulst} which controls either the birth rate or external death rate. The bit string
is $32/64$ bits long to enable bitwise manipulation on integer type variables. The finite length bit string
is an artifice of simulation and is not an important consideration when approaching the model analytically.
A solution to the Penna model
in the steady-state has been developed and is capable of dealing with a wide range of modifications
to the model, namely arbitrary birth and survivability functions \cite{Analytical_solution}.

By building upon the steady-state solution 
it is possible to consider dynamic behaviour. The simplest form the steady-state solution takes is for the simple 
Penna model in which there is no non-genetic source of death, an organism dies after a single 
disease ($T=1$) and can reproduce with equal probability at any point during its life. For simplicity
we present our dynamics analysis within this simple Penna model, since it is straightforward to generalise
to the case of $T>1$ \cite{Analytical_solution}. The steady-state solution \cite{Coe_Mao_Cates} 
to the simple Penna model is given in brief below:

An organism within the Penna model can be uniquely characterised by its age $x$ and the number of $0$s
on its bit-string before the first $1$. The number of $0$s determines how long the organism lives
and is termed its string-length $l$.
Where $n_j(x,l)$ is the number of organisms with age $x$ and 
string-length $l$ at time-step $j$ 
\begin{eqnarray}
n_{j+1}(0,l)&=&be^{-\beta l} \sum_{x=0}^\infty n_j(x,l)\label{Penna_Equation}\\ 
&+& mbe^{-\beta l}\sum_{l'>l}^\infty \sum_{x=0}^\infty n_j(x,l'),\nonumber
\end{eqnarray}  
where $e^{-\beta}=1-m$ is the probability of avoiding a mutation. 
In the stationary-state $n_{j+1}(x,l)=n_j(x,l)$ and given that
an organism with string-length $l$ lives for $l$ timesteps, the sum over ages
of $n(x,l)$ is $l\times n(0,l)$. The sum over ages is written as $n(l)=\sum_x n(x,l)$.
Then eqn.\ (\ref{Penna_Equation}) simplifies to
\begin{equation}
0=be^{-\beta l} n(l) 
- \frac{n(l)}{l}
+ mbe^{-\beta l}\sum_{l'>l}^\infty n(l').
\end{equation}  
This expression can be solved to generate a recursion relation
\begin{equation}
\frac{n(l+1)}{n(l)}=\frac{l+1}{l}\frac{ e^{\beta l} - bl }
{ e^{\beta(l+1)} - b(l+1)e^{-\beta}}.\label{SPM_solution}
\end{equation}
For any population there is a maximum sustainable string-length $l_{\mathrm{max}}$, 
the stability analysis of which \cite{Analytical_solution, Coe_Mao_Cates}
leads the stationary-state interdependence of $b$ and $\beta$:
\begin{eqnarray}
l_{\mathrm{max}}&<&\frac{1}{1-e^{-\beta}}.\label{SPM_Lmax}\\
b&=&\frac{1}{l_{\mathrm{max}}}e^{\beta l_{\mathrm{max}}}.\label{SPM_b}
\end{eqnarray}  
For finite-length bit-strings, $l_{\mathrm{max}}$ of course cannot exceed the bit string itself.
\begin{figure}
 \includegraphics[width=3in]{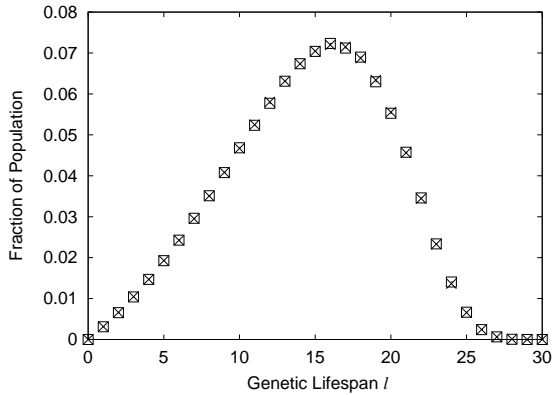}
 \caption{\label{fig1}Lifespan distribution for a simple Penna model with $l_{\mathrm{max}}=30$,
$\beta=\frac{1}{30}$. Analytical results ($\times$) are compared with those from simulation ($\square$
). Simulation size $10^7$.}
\end{figure}


\section{Dynamics in the Simple Penna Model}

Penna model dynamics can be divided into three cases: (A) The birth rate or/and mutation rate are altered from 
their steady state values leading to growth or decline of the total population $n$, and associated changes 
in the distribution $n(l)$ of the sub-populations with string-lengths $l$. (B) A change in $n(l)$
from the steady-state distribution predicted in equation (\ref{SPM_solution}) is followed by relaxation back
to the steady-state. 
(C) Within a sub-population $n(l)$, the distribution of ages $n(x,l)$ can fluctuate with time
if the population is not in steady state.


\subsection{Steady Growth and Decline in the Simple Penna Model}
With an unsuitable choice of birth rate, mutation rate and maximum string-length present in a population,
 stationary-state behaviour will not be found. Eventually the behaviour of the entire population will be 
dominated by the growth or decline of the longest $l$-type.
The population can exist in a state of steady growth or decline in which the relative sizes of sub-populations
 remain the same. The governing equation can be written as
\begin{eqnarray}
n_{j+1}(0,l)&=&be^{-\beta l} \sum_{x=0}^\infty n_j(x,l)\\ 
&+& mbe^{-\beta l}\sum_{l'>l}^\infty \sum_{x=0}^\infty n_j(x,l').\nonumber
\end{eqnarray}  
If we label the rate of growth $r$, then in any timestep
the number of young produced is $1+r$ times that in the previous timestep.
Once the population has been growing in this manner for some time, populations at successive iterations
are related by $n_{j+1}(x,l)=(1+r)n_j(x,l)$.
As in the steady-state case, $n(l)$ is defined to be the sum over ages of $n(x,l)$.
We define $L_r(l)$ so that
$n(l)=L_r(l)n(0,l)$. This leads to a simplified steady-growth equation:
\begin{eqnarray}
0 &=&be^{-\beta l} n_j(l) - \frac{(1+r)}{L_r(l)}n(l)\\ 
&+& mbe^{-\beta l}\sum_{l'>l}^\infty n_j(l').\nonumber
\end{eqnarray}  
Where $L_r(l)$ is given by
\begin{eqnarray}
L_r(l)=\frac{1-\Big(\frac{1}{1+r}\Big)^l}{1-\frac{1}{1+r}}\label{L_r}
\end{eqnarray}
The steady growth equation can be manipulated to give a recursion relation for the
relative sizes of successive $n_j(l)$
\begin{eqnarray}
\frac{n_j(l+1)}{n_j(l)}&=&\frac{L_r(l+1)}{L_r(l)} \label{steady-growth}\\
&\times&\frac{ (1+r)e^{\beta l} - bL_r(l) }
{ (1+r)e^{\beta(l+1)} - bL_r(l+1)e^{-\beta}}.\nonumber
\end{eqnarray}

\begin{figure}
 \includegraphics[width=3in]{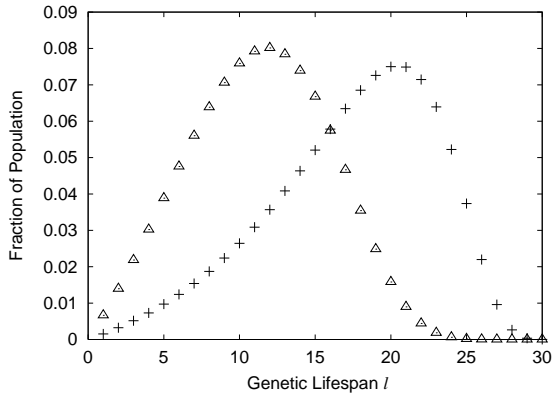}
 \caption{\label{fig2}Lifespan distributions for a Penna model undergoing steady growth with $l_{\mathrm{max}}=30$,
$r=0.05$ ($+$) and $l_{\mathrm{max}}=30$, $r=-0.05$ ($\triangle$).}
\end{figure}

The conditions for steady-growth give a limit on the value of the maximum sustainable string-length
$l_{\mathrm{max}}$ and determine the value of the birth rate $b$.
\begin{eqnarray}
l_{\mathrm{max}}&<&
\frac{
\ln\Big(\frac{1-e^\beta}{1-e^\beta(1+r)}\Big)
}{\ln \Big(\frac{1}{1+r}\Big)}\\
b&=&\frac{re^{\beta l_{\mathrm{max}}}}{1-\Big(\frac{1}{1+r}\Big)^{l_{\mathrm{max}}}}
\end{eqnarray}
In the limit of vanishing $r$, a power series expansion of these expressions will give, to leading order,
equations (\ref{SPM_Lmax},\ref{SPM_b}).

\subsection{Sub-population Dynamics in the Simple Penna Model} 

For sub-population dynamics we consider the time-step evolution of an arbitrary distribution of
$n(l)$ ($l\leq l_{\mathrm{max}}$) where the birth and mutation rate take their steady-state values.
The dynamics of sub-populations within the Penna model can be considered to be that of a series of decay modes.
Any sub-population can be expressed as the sum over contributions from a set of decay modes such that
\begin{eqnarray}
n_j(x,l)=\sum_k A_k n_{k,j}(x,l),
\end{eqnarray}
where the constants $A_k$ are to be determined.
Within each decay mode, labelled by its index $k$, the population dies away exponentially so that
\begin{eqnarray}
n_{k,j+1}(x,l)=(1-\lambda_k) n_{k,j}(x,l).\label{lambda}
\end{eqnarray}
Within each mode there is a maximum value of $l$ above which $n_{k,j}(l)$ is zero, this is labelled $l_k$.
The mode index $k$ is chosen so that $k$ is the number of non-zero sub-populations within the mode.
Once the relationship between $n_j(x,l)$ and $n_{j+1}(x,l)$ has been established
equation (\ref{Penna_Equation}) can be considered to be a sum of eigen-equations
each of which governs the behaviour of a given mode.
The equation governing the time-step evolution of the longest string-length within a mode can be written as
\begin{eqnarray}
n_{k,j+1}(0,l_k)&=&be^{-\beta l_k} \sum_{x=0}^\infty n_{k,j}(x,l_k).
\end{eqnarray}
The time step evolution of $n_{k,j}(x,l)$ is known (\ref{lambda}) so the sum over ages can be evaluated and 
$\lambda_k$ identified as the solution to the equation
\begin{eqnarray}
\lambda_k=1-\frac{1-\Big(\frac{1}{1-\lambda_k}\Big)^{l_k}}{1-\frac{1}{1-\lambda_k}}be^{-\beta l_k}.
\end{eqnarray}
The characterstic decay time $\tau_k$ for any mode is defined as the time taken for the mode to decay to
$e^{-1}$ of its initial size.

\begin{figure}
 \includegraphics[width=3in]{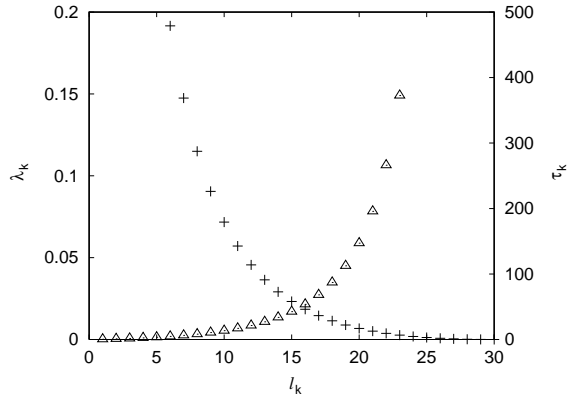}
 \caption{\label{fig3}$\lambda_k$ (+) and decay time $\tau_k$ ($\triangle$) plotted against $l_k$ 
for a simple Penna model with $l_{\mathrm{max}}=30$, $\beta=\frac{1}{30}$.}
\end{figure}

Manipulation of equation (\ref{Penna_Equation}) for an individual mode gives a recursion relation
for the relative sizes of sub-populations within a mode
\begin{eqnarray}
\frac{n_{k,j}(l+1)}{n_{k,j}(l)}&=&\frac{L_k(l+1)}{L_k(l)}\\&\times&\frac{ (1-\lambda_k)e^{\beta l} - bL_k(l) }
{ (1-\lambda_k)e^{\beta(l+1)} - bL_k(l+1)e^{-\beta}},\nonumber
\end{eqnarray}
where $L_k(l)$ is given by
\begin{eqnarray}
L_k(l)=\frac{1-\Big(\frac{1}{1-\lambda_k}\Big)^l}{1-\frac{1}{1-\lambda_k}}\label{L_k}.
\end{eqnarray}
Each mode mimics the behaviour of a population undergoing steady decline, as given in equation 
(\ref{steady-growth}). 
Note that the dynamic nature of this model means that the general steady-state solution presented in 
\cite{Analytical_solution}
does not give a recursion relation of exactly this nature. Naturally in the limit of vanishing $\lambda_k$
the recursion relation above gives that from the steady state Penna model.

\begin{figure}
 \includegraphics[width=3in]{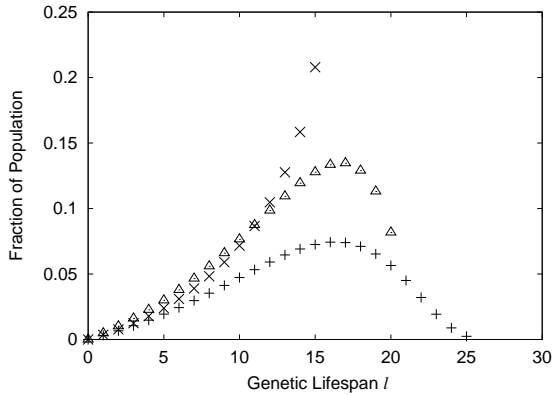}
 \caption{\label{fig4}
Sizes of sub-populations within decay modes for $\beta=\frac{1}{30}$, $l_{\mathrm{max}}=30$ with $l_k=25$ ($+$), 
$l_k=20$ ($\triangle$) and $l_k=15$ ($\times$). These plots have been rescaled (they are not all normalized to $1$) 
to plot them all on the same axes.} 
\end{figure}

$n_{k,j}(l)$ retains its meaning from the steady-state analysis as a sum of $n_{k,j}(x,l)$ over ages $x$.
As in the steady state analysis, $L_k(l)$ gives the sum over ages: $n_{k,j}(l)=L_k(l)n_{k,j}(0,l)$
The decay mode for which $l_k=l_{\mathrm{max}}$ has a decay rate of $0$ ($\lambda_k=0$) and is the
steady-state solution.

An arbitrary distribution of $n_j(l)$ can be broken down into a sum over decay modes. 
Decomposition of an $n_j(l)$ distribution into decay modes can be done using a `top down' 
approach: The largest value of $l$ for which $n_j(l)\neq 0$ gives the $l_k$ value for the largest mode, $A_k$ is
then chosen so that $n_j(l_k)=A_k n_{k,j}(l_k)$. Having determined $A_k$, $A_{k-1}$ follows as:
$n_j(l_k-1)=A_k n_{k,j}(l_k-1)+A_{k-1}n_{k-1,j}(l_k-1)$. This process is repeated,
accounting for all contributions from higher modes at each $l$, until all $A_k$ are determined. 
Any distribution of $n_j(l)$ ($n(l)=0$ for $l>l_{\mathrm{max}}$) can be uniquely broken down into decay modes.

\subsection{Age Distribution Dynamics in the Simple Penna Model}
The dynamic and steady-state behaviour of the simple Penna model is dependent on the distribution of ages 
taking a particular form. An arbitrary distribution of $n(x,l)$ will have its own dynamic behaviour and must be 
considered as a seperate case. In the simple Penna model the evolution of $n(x,l)$ over time without 
contributions from mutation can be dealt with
by a Leslie Matrix \cite{Leslie_matrices} of rank $l$ acting on a vector $n(x,l)$ where the vector 
components correspond to ages. For $n(x,l)$ the allowed ages are $0,1,2,3\dots l-1$ and the timestep evolution 
can be descibed by the following matrix equation
\begin{eqnarray}
\left( \begin{array}{c}
n(0,l)\\
n(1,l)\\
n(2,l)\\
n(3,l)\\
\vdots\\
\end{array} \right)_{j+1}&=\boldsymbol{A}
\left( \begin{array}{c}
n(0,l)\\
n(1,l)\\
n(2,l)\\
n(3,l)\\
\vdots\\
\end{array} \right)_{j}
\end{eqnarray}
where the matrix $\boldsymbol{A}$ is given by
\begin{eqnarray}
\left( \begin{array}{ccccc}
be^{-\beta l}&be^{-\beta l}&be^{-\beta l}&be^{-\beta l}&\dots\\
1&0&0&0&\dots\\
0&1&0&0&\dots\\
0&0&1&0&\dots\\
\vdots&\vdots&\vdots&\vdots&\ddots\\
\end{array} \right)
\end{eqnarray}
We require eigenvalue solutions to this so that
\begin{eqnarray}
n(x,l)_{j+1}=\xi n(x,l)_{j}
\end{eqnarray}
$\xi$ are the eigenvalues of the matrix $\boldsymbol{A}$. 
These eigenvalues can be identified as non-unity solutions to the polynomial
\begin{eqnarray}
\xi^{l+1}-(be^{-\beta l}+1)\xi^{l}+be^{-\beta l}=0.
\end{eqnarray}
The large eigenvalues give decay at roughly the rate predicted from the $n(l)$ analysis and oscillation
around this decay. The small eigenvalues play no significant part over large time scales. 
$\xi_{\mathrm{max}}$ the largest eigenvalue, is related to the decay rate of the decay mode $n_{kj}(l)$
by $\lambda_k = 1 - \xi_{\mathrm{max}}$. If $l<<\frac{1}{\lambda_k}$ then the long term behaviour of any 
perturbation
in $n(l)$ will be dictated by a series of decay modes.

\section{Analysis of Computational Dynamics }
A Simple Penna model simulation was run with $b$ and $\beta$ chosen to give $l_{\mathrm{max}}=30$.
The population was initialised with $n(x,l)=0$ with the exception of a spike at $n(0,25)$ where
$10^5$ organisms were created. The simulation was then allowed to run generating data for $n(l)$ at each
iteration. As the maximum lifespan in the population is less than $l_{\mathrm{max}}$ the population will eventually
decay away to nothing. The spike was chosen as the intial distribution as any initial distribution can be 
considered to be a sum of such spikes.

\begin{figure}
\includegraphics[width=3in]{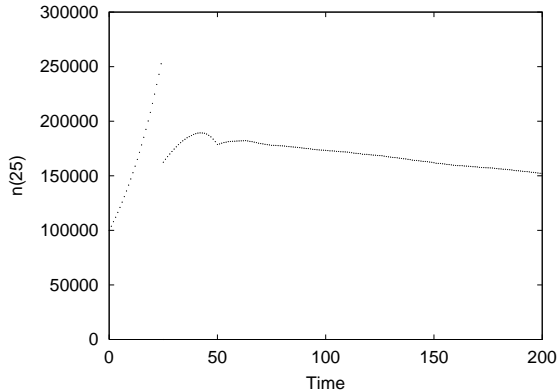}
\caption{\label{n25_plot}A plot of n(25) against time from simulation results. The population is initialised with 
$n(0,25)=10^5$ and all other $n(x,l)$ set to zero. The sudden drop in population at $25$ timesteps 
arises because all of the initial population die once they reach age $25$. After $100$ iterations the noise 
induced by age distribution dynamics has essentially disappeared and the population dies away exponentially 
with time constant $\tau_k$ of $832.4$.}
\end{figure}

\begin{figure}
\includegraphics[width=3in]{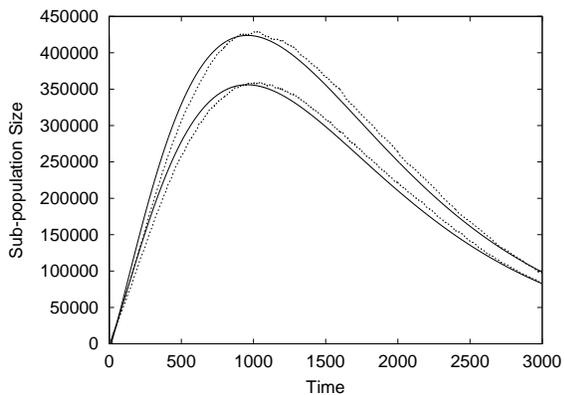}
\caption{\label{nDecay_plot}Analytical results for the decay of $n(20)$ and $n(10)$ 
are compared with those from simulation. The upper graphs correspond to $n(20)$,
the lower to $n(10)$. Analytical results are given by the solid lines, simulation results by the dotted lines.
The analytical results are obtained using only the highest 4 modes.}
\end{figure}

Initially the dynamics of $n(25)$ will be dominated by age distribution dynamics and cannot be 
explained in terms of modes. After $100$ iterations this noise has all but disappeared and the subpopulation can 
be seen to decay exponentially as predicted by equation ({\ref{lambda}}) (see fig.~\ref{n25_plot}). 
All other subpopulations initially grow to later enter a period of exponential decay (see fig~\ref{nDecay_plot}). 

Considering the population after $100$ iterations, modes 
with $\tau_k<100$ will play an insignificant part in the long term dynamics of any subpopulation. After large 
number of timesteps $t$, each sub-population, to a good approximation, can be represented by the highest few 
modes (with largest decay times):
\begin{eqnarray}
n_t(l)&\approx&A_{25}n_{25,0}(l)e^{\frac{-t}{\tau_{25}}}\\
&+&A_{24}n_{24,0}(l)e^{\frac{-t}{\tau_{24}}}\nonumber\\
&+&A_{23}n_{23,0}(l)e^{\frac{-t}{\tau_{23}}}\nonumber\\
&+&A_{22}n_{22,0}(l)e^{\frac{-t}{\tau_{22}}}\nonumber ...
\end{eqnarray}
The notation used is $n_{k,t}(l)$ where $k$ is the mode index, 
$t$ is the number of timesteps after the $100$th iteration.
$A_k$ are mode coefficients (determined from the distribution of $n(l)$ at the $100th$ iteration).

Taking more modes into account will give a more accurate picture so that the behaviour of any subpopulation can
be described exactly (in the absence of age dynamic induced noise) by
\begin{eqnarray}
n_t(l)&=&\sum_{k=1}^{25}A_{k}n_{k,0}(l)e^{\frac{-t}{\tau_{k}}}.
\end{eqnarray}

\section{Conclusion}
As a natural extension to our work on the steady-state Penna model we have considered and obtained analytical 
expressions for the various forms of dynamics which can be displayed by a simple Penna model.
Our approach can be applied to the governing equations derived in \cite{Analytical_solution} to analyze the
dynamics of a variety of Penna models, such as those with external death rates and birth cutoffs.
For multiple disease Penna models ($T>1$) the breakdown of the population into decay 
modes remains valid so long as there is no bias in the
distribution of non-terminal mutations within bit-strings.

The decay coefficients ($\lambda_k$) and
relationship between $n_{k,j}(l)$ within modes for models with $T>1$.
are given as they may be of particular interest.
Each $\lambda_k$ satisfies
\begin{eqnarray}
\lambda_k=1-\frac{1-\Big(\frac{1}{1-\lambda_k}\Big)^{l_k}}{1-\frac{1}{1-\lambda_k}}be^{-\beta(l_k-T+1)}.
\end{eqnarray}
The recursion relation between successive sub-populations within a mode is
\begin{eqnarray}
\frac{n_{k,j}(l+1)}{n_{k,j}(l)}&=&\frac{C^{l+1}_{T-1}}{C^{l}_{T-1}}\frac{L_k(l+1)}{L_k(l)}\\
&\times&\frac{ (1-\lambda_k)e^{\beta(l-T+1)} - bL_k(l) }
{ (1-\lambda_k)e^{\beta(l+1-T+1)} - bL_k(l+1)e^{-\beta}}.\nonumber
\end{eqnarray}
where $L_k(l)$ is unchanged from the single mutation case and given by equation (\ref{L_k}).

The dynamic behaviour we have considered is for unregulated populations where there is no Verhulst factor
and steady state is obtained by suitable choice of birth rate and mutation rate. 
Dynamics in the presence of a Verhulst factor is considerably more complicated as any change in a sub-population
will act to change the birth (or death) rate, affecting all other sub-populations. 
The modes, which in the absence of a Verhulst factor are independent, become coupled through a shared birth (or 
death) rate.
The resulting series of coupled, non-linear difference equations cannot be generally
studied using the decay modes analysis
we have presented. However, depending how the Verhulst factor is implemented there may be regimes in which 
the coupling between decay modes is sufficiently weak that they may be treated as independent. 

The authors would like to thank M.~E.~Cates and R.~Haydock for useful advice and discussion.

\end{document}